\documentclass[journal=nalefd,manuscript=article]{achemso}

\usepackage[version=3]{mhchem} 
\usepackage{epstopdf}
\usepackage{amsmath}
\usepackage{amssymb}
\usepackage{dcolumn}
\usepackage{graphicx}
\usepackage{longtable}
\usepackage{bm}
\usepackage{color}
\usepackage{epsfig}
\usepackage{hyperref}
\usepackage{bookmark}

\author{Madhav Prasad Ghimire}
\affiliation[Leibniz Institute for Solid State and Materials Research, IFW Dresden]
{Leibniz Institute for Solid State and Materials Research, IFW Dresden, Helmholtzstr. 20, D-01069 Dresden, Germany}
\alsoaffiliation[Condensed Matter Physics Research Center (CMPRC)]
{Condensed Matter Physics Research Center (CMPRC), Butwal-11, Rupandehi, Nepal}
\email{m.p.ghimire@ifw-dresden.de}
\author{Manuel Richter}
\affiliation[Leibniz Institute for Solid State and Materials Research, IFW Dresden]
{Leibniz Institute for Solid State and Materials Research, IFW Dresden, Helmholtzstr. 20, D-01069 Dresden, Germany}
\alsoaffiliation[Dresden Center for Computational Materials Science (DCMS)]
{Dresden Center for Computational Materials Science (DCMS), TU Dresden, D-01062 Dresden, Germany}
\email{m.richter@ifw-dresden.de}
\phone{+49 (0)351 4659360}
\fax{+49 (0)351 4659750}

\title[An \textsf{achemso} demo]
  {Chemical Gating of a Weak Topological Insulator: Bi$_{14}$Rh$_3$I$_9$}

\begin{document}


\begin{abstract}
The compound Bi$_{14}$Rh$_3$I$_9$ has recently been suggested as a weak 3D topological insulator (TI) on the basis of angle-resolved photoemission and
scanning-tunneling experiments in combination with density-functional (DF) electronic structure calculations.
These methods unanimously support the topological character of the headline compound, but a compelling confirmation could only be obtained by dedicated transport experiments.
The latter, however, are biased by an intrinsic n-doping of the material's surface due to its polarity.
Electronic reconstruction of the polar surface shifts the topological gap below the Fermi energy, which would also prevent any future device application.
Here, we report the results of DF slab calculations for chemically gated and counter-doped surfaces of Bi$_{14}$Rh$_3$I$_9$.
We demonstrate that both methods can be used to compensate the surface polarity without closing the electronic gap.

\end{abstract}

\subsection{Keywords:} Bi$_{14}$Rh$_3$I$_9$, topological insulators, density functional theory, chemical gating, doping, quantum spin Hall effect\\\

In recent years, topological insulators (TIs) have attracted enormous attention due to their intriguing properties.
Of particular interest are their massless Dirac-cone-like surface states protected by time-reversal symmetry (TRS)~\cite{zhang,chen,hasan}.
In a nutshell, TIs are characterized by these gapless surface states and a bulk energy gap.

Three dimensional (3D) TIs are called strong or weak based on four $Z_2$ invariants ($\nu_0;\nu_1,\nu_2,\nu_3)$.
If $\nu_0 = 1$, the material is a strong TI; if $\nu_0 = 0$ and any of the indices $(\nu_1,\nu_2,\nu_3)$ is equal to one, it is a weak TI ~\cite{fu,moore}.
In the former case, including the well-known compounds Bi$_2$Se$_3$ and Bi$_2$Te$_3$~\cite{zhang,chen}, the TRS-protected surface states are present on all facets, 
while in the latter case, such surface states are present only on certain facets.
Their peculiar properties bear the potential to enable novel types of information processing~\cite{moore2010,brune2012,alicea2012}. 
An important step toward so-called topological quantum computing is the recent observation of signatures of Majorana fermion modes~\cite{qinglinhe2017}. 
In closer reach could be an application of the recently observed excellent thermoelectric properties of the title compound ~\cite{wei2016}
or of the all-electrical detection of spin polarization by combination 
of quantum spin Hall and metallic spin Hall transport in a single device~\cite{brune2012}.

Weak 3D TIs suggested hitherto are usually hosted by layered crystal structures~\cite{reviewbansil}.
The strength of the related inter-layer coupling influences the bulk band structure:
(i) In Bi$_2$TeI with strong inter-layer coupling, this coupling is essential for the formation of the weak 3D TI state \cite{tang};
(ii) a weak interlayer coupling, on the other hand, results in a quasi two-dimensional (2D) band structure.
This situation is found in KHgSb \cite{yan}, ZrTe$_5$ \cite{wu} or Bi$_{14}$Rh$_3$I$_9$ \cite{rasche,pauly1}.
Weak 3D TIs of the second kind may allow to produce 2D TI structures that are expected to show the quantum spin-Hall (QSH) effect.
This can be achieved by cleaving off thin layers from the bulk 3D TI as an alternative way to the fabrication of quantum wells \cite{slager,weng}.
Indeed, a single, charge-compensated layer of Bi$_{14}$Rh$_3$I$_9$ was predicted to be a 2D TI in a recent calculation \cite{rasche2016}.

\begin{figure}
     \centering
     \includegraphics[width=0.68\textwidth]{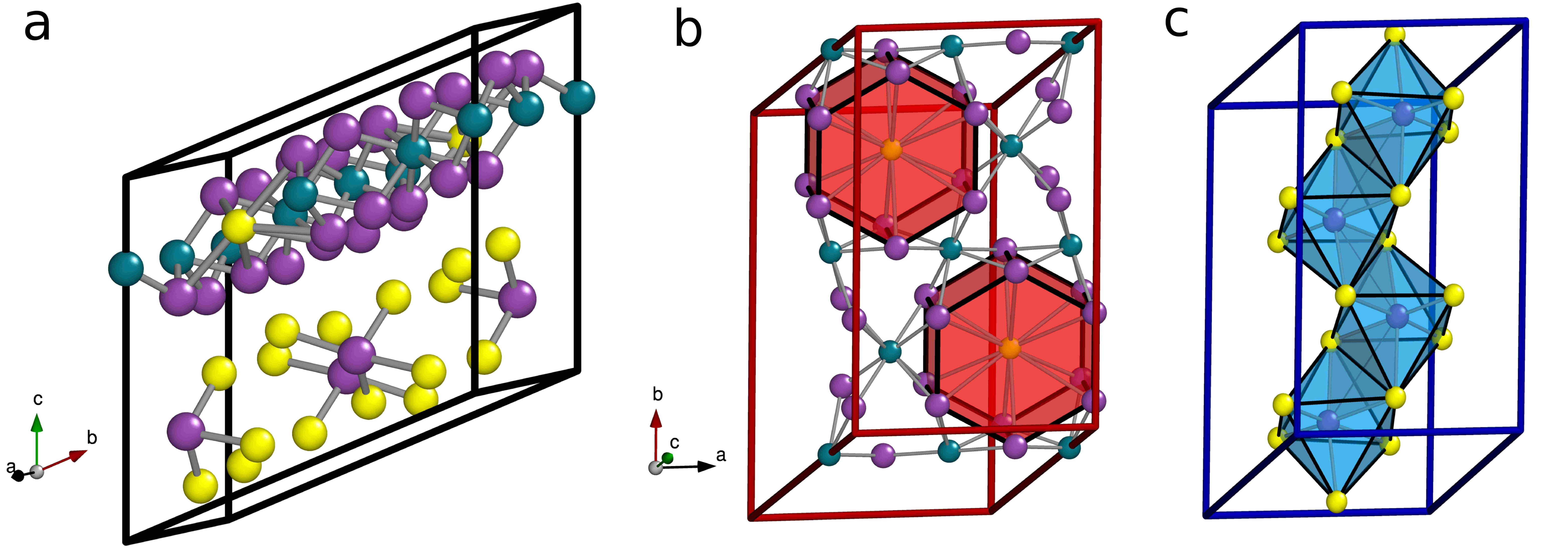}
     \caption{Bulk structure of Bi$_{14}$Rh$_3$I$_9$. (a) Elementary cell with [(Bi$_4$Rh)$_3$I]$^{2+}$ in the upper part and [Bi$_2$I$_8$]$^{2-}$ in the lower part.
              The color code is violet for Bismuth, green for Rhodium and yellow for Iodine.
              (b) Rotated view of the [(Bi$_4$Rh)$_3$I]$^{2+}$ layer, also denoted as 2D TI layer;
              (c) the same for the [Bi$_2$I$_8$]$^{2-}$ layer, also denoted as spacer layer.
              The red/blue bounding boxes in (b)/(c) establish a correspondence to the schematic structures shown in Figure~\ref{fig:struc}.}
     \label{fig:structure}
 \end{figure}

The recently synthesized compound Bi$_{14}$Rh$_3$I$_9$ was characterized as a layered ionic structure with alternating cationic [(Bi$_4$Rh)$_3$I]$^{2+}$ and anionic [Bi$_2$I$_8$]$^{2-}$ layers, as shown in Figure~\ref{fig:structure}.
Density functional (DF) calculations for this material found ($\nu_0;\nu_1,\nu_2,\nu_3) = (0;0,0,1)$.
Further, the electronic band-structure grossly agreed with angle-resolved photoemission spectra (ARPES) obtained on single crystals.
On this basis, Bi$_{14}$Rh$_3$I$_9$ was claimed to be a weak TI \cite{rasche}.

Subsequently, this hypothesis was strengthened by scanning tunneling microscopy (STM) experiments \cite{pauly1}.
By STM topography, the investigated [001] surface was found to exhibit areas with both types of layers.
Clear signatures of one-dimensional (1D) states were observed in the band gap only at step edges of cationic surface layers.
However, the related surface-layer gap was found 0.25 eV below the Fermi level ($E_{\text{F}}$) \cite{pauly1}.
The [(Bi$_4$Rh)$_3$I]$^{2+}$ layer carrying the edge states was found to be structurally intact.
The [Bi$_2$I$_8$]$^{2-}$ layer, on the other hand, contained holes which were attributed to the evaporation of Iodine atoms during cleavage.
Such a {\em chemical reconstruction} is one possibility~\cite{noguera} to compensate the obvious surface polarity of the system.
DF calculations confirmed the observed down-shift of the topologically nontrivial band gap at the cationic [001] surface \cite{pauly2}.
This is a clear sign of an {\em electronic reconstruction} as a second possibility~\cite{noguera} to compensate surface polarity.

A confirmation of the weak 3D TI state of Bi$_{14}$Rh$_3$I$_9$ would require to observe the QSH effect on the mentioned 1D edge states~\cite{chengliu}.
However, related transport experiments make only sense if the observed intrinsic doping is compensated by reasonable means and, thus, the topological gap with the edge states is shifted to the Fermi level.

There are several possible ways to compensate the surface polarity:
(i) physical gating by preparation of a dielectric gate structure and applying the electric field effect;
(ii) chemical gating by deposition of an oxidizing agent; or
(iii) counter-doping of the surface layer.
Here, we report results of investigations into the two latter possibilities by means of DF calculations.
In particular, we study the effects of Iodine deposition as a sparse overlayer and of counter-doping by exchanging surface-layer Bi atoms by Sn.
The results are expected to provide suggestions for the preparation of forthcoming transport experiments, which are required to confirm the topological state of Bi$_{14}$Rh$_3$I$_9$ or similar systems.
We are not aware of any observation of the QSH effect on another suggested weak 3D TI material.

All DF calculations were done with the full-potential local-orbital (FPLO) code \cite{klaus} using the local density approximation (LDA) in the PW92 parametrization \cite{perdew}.
The self-consistent calculations were carried out with spin-orbit coupling (SOC) included.
This effort is necessary because the involved elements have a sizable SOC strength which is responsible for opening the band gap.
The following basis states are treated as valence states: Bi: 5s, 5p, 5d, 6s, 7s, 6p, 7p, 6d; Rh: 4s, 4p, 5s, 6s, 4d, 5d, 5p; I: 4s, 4p, 4d, 5s, 6s, 5p, 6p, 5d.

In order to simulate the [001] surface of a bulk sample, we considered a series of slabs with different thickness (1.25 to 3.75 nm), varying from one to three structural layers.
The considered layer stacks have the same lateral cell dimensions as the experimental bulk structure \cite{rasche}, and equivalent atomic positions.
Thus, the elementary cell of a slab with one, two, and three structural layers contains two, four, and six chemical units, respectively.
The slabs were treated in a 3D supercell geometry along [001] with vacuum layers of 2.50 nm thickness.
The $k-$space integrations were carried out with the linear tetrahedron method using a 6 x 4 x 1 $k-$mesh in the full Brillouin zone.

\begin{figure}
      \centering
      \includegraphics[width=0.4\textwidth]{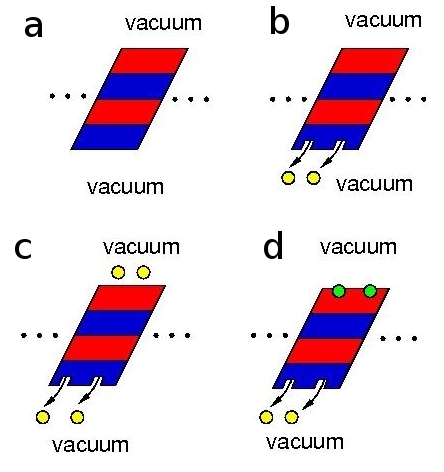}
      \caption{Schematic structural models. Slabs with two structural layers are depicted as examples.
 Blue (red) sections of the schemes indicate the spacer (2D TI) layers. Yellow balls indicate Iodine atoms, and green balls indicate Tin atoms. 
(a) Unreconstructed slab; (b) Chemically reconstructed slab (desorption of Iodine atoms from the spacer surface);
 (c) Adsorption of Iodine atoms on top of the 2D TI surface; (d) Chemical doping by Tin within the topmost atomic layer of the 2D TI surface.}
      \label{fig:struc}
 \end{figure}

As mentioned before, Bi$_{14}$Rh$_3$I$_9$ consists of alternating layers:
a cationic [(Bi$_4$Rh)$_3$I]$^{2+}$ layer with a nontrivial gap~\cite{rasche2016} and an
anionic [Bi$_2$I$_8$]$^{2-}$ layer with a trivial gap~\cite{pauly2} (see Figure~\ref{fig:structure}).
The former one is denoted as 2D TI layer, the latter is denoted as spacer layer.
Figure~\ref{fig:struc} shows schematic slabs with two structural layers.
These slabs contain both possible surface types, cationic and anionic.
The advantages of this choice consist in a correct stoichiometry and in the possibility to
study the electronic structure of both surface types in one and the same calculation.
However, the chosen structure model does not obey inversion symmetry (space group 1).
Further, slabs with an even number of 2D TI layers are expected to be topologically trivial~\cite{imura2012}.
Thus, the structure shown in Figure~\ref{fig:struc}(a) cannot be considered as a possible model for a 2D TI, but it is rather used to study the electronic structure.
In a realistic structure, however, step edges ensure that there is a large spatial distance between edge states of neighboring 2D TI layers, in addition to the weak interlayer coupling.
Thus, only an exponentially small edge state gap is expected.

We consider chemical modifications on both surfaces.
First, the experimentally observed desorption of Iodine from the spacer layer is modelled,
where about two Iodine atoms per surface elementary cell are removed during cleavage (see Figure~\ref{fig:struc}(b))~\cite{pauly1,pauly2}.
Earlier, the Iodine desorption was investigated by means of a DF virtual crystal approximation (VCA) approach~\cite{pauly2},
where the additional charge is equally distributed among all atoms of the outermost atomic layer.
Here, we apply a more realistic structural model with specific Iodine atoms removed from that layer.
Second, adsorption of a sparse Iodine layer on top of the 2D TI surface, see Figure~\ref{fig:struc}(c), is investigated for the sake of tuning $E_{\text{F}}$.
Third, we investigate the effect of surface doping by replacing part of the Bi atoms in the outermost atomic layers of the 2D TI surface by Sn, Figure~\ref{fig:struc}(d).
Details of all structural models used are given in the Supplementary Information.

\begin{figure}
      \centering
      \includegraphics[width=0.68\textwidth]{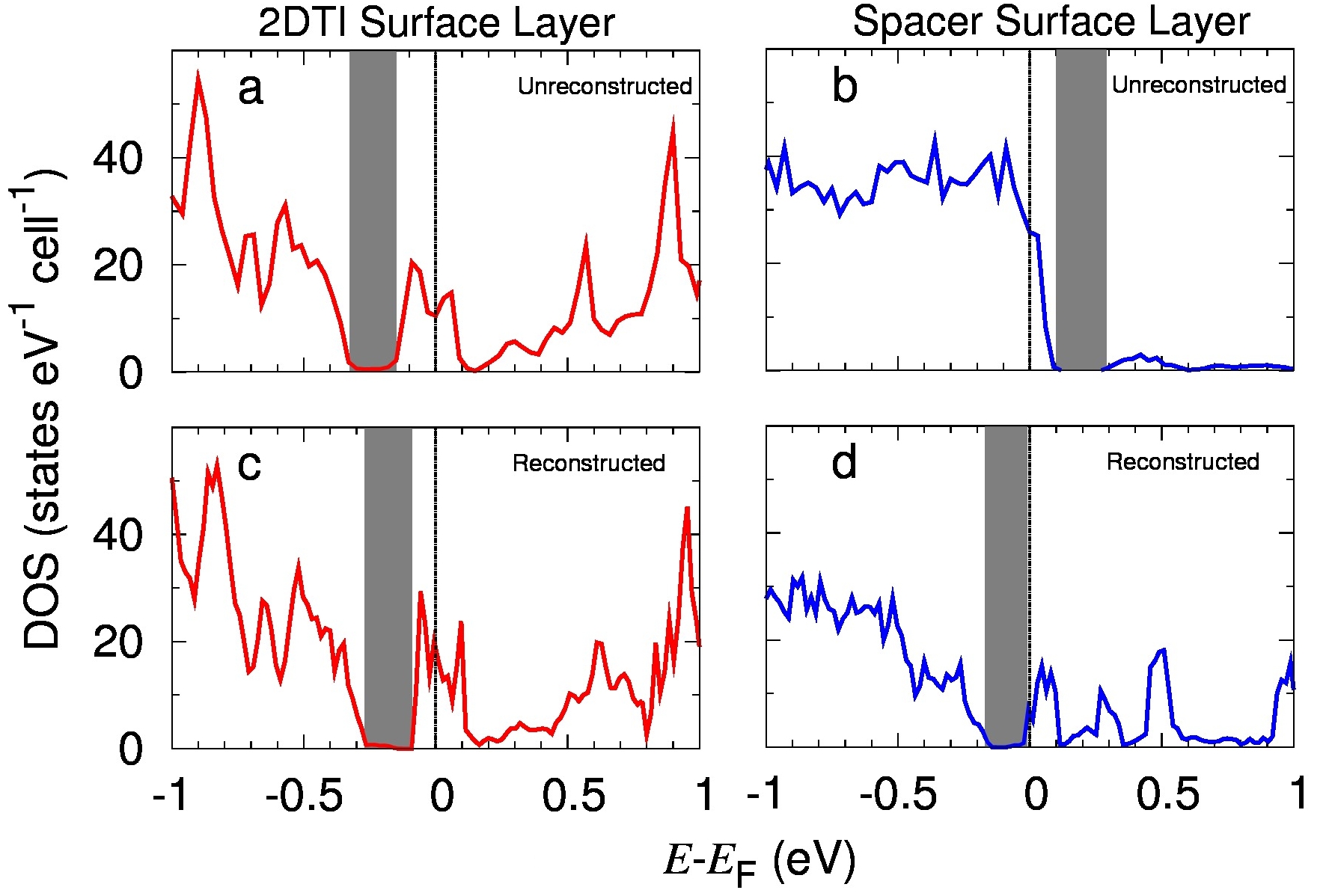}
      \caption{Layer-resolved density of states (DOS) of 2D TI surface layers (Bi$_4$Rh)$_3$I (a, c) and spacer surface layers Bi$_2$I$_8$ (b) and
               Bi$_2$I$_6$ (d) of slabs with two structural layers.
               Unreconstructed (a, b) and chemically reconstructed (c, d) systems are shown.
               In the reconstructed system, two Iodine atoms per surface unit cell are removed from the spacer surface layer.
               Note, that the data of (b) has already been presented in Ref.~\cite{pauly2}.
               It is included here to have the present paper sufficiently self-contained.}
      \label{fig:desorption}
 \end{figure}

\textbf{\emph{Desorption of Iodine from the spacer surface:}}
We performed DF calculations for two-layer slabs without (Figure~\ref{fig:struc}(a)) and with (Figure~\ref{fig:struc}(b)) chemical
reconstruction of the anionic spacer surface in order to compare with experimental information and with the previous VCA results \cite{pauly2}.
Figure~\ref{fig:desorption} shows the surface-layer projected density of states (DOS) for both cases.
We first consider the case of an unreconstructed spacer surface.
As observed in Figure~\ref{fig:desorption}(a), a band gap of about 0.18 eV, centered at about -0.25 eV, is formed at the 2D TI surface.
The surface spacer DOS at $E_{\text{F}}$ is high, Figure~\ref{fig:desorption}(b), and dominated by Iodine-5p states.
The large DOS indicates that the unreconstructed structure could be unstable.
The 2D TI surface DOS which has predominant Bi-6p character is in good agreement with spectroscopic data from ARPES~\cite{rasche} and STS~\cite{pauly2},
which both show a gap of about 0.15 eV width at about -0.25 eV. On the other hand, the spacer surface DOS shows no gap
below $E_{\text{F}}$ and clearly contradicts the quoted experimental findings.

Recent calculations simulated the observed Iodine deficiency~\cite{pauly1} by means of VCA, thus distributing the remaining
charge among all Iodine sites of the outermost atomic layer of the spacer surface. These calculations reproduced
the observed gap on the spacer surface~\cite{pauly2}.
However, no atomistic calculation was performed, hitherto, to confirm the reliability of the VCA results.
Here, we present atomistic DF results for the chemically reconstructed spacer surface where two Iodine atoms are removed per surface unit cell 
(SUC) with an area of 1.452$\times 10^{-18}$ m$^2$, Figure~\ref{fig:desorption}(c, d).
From Figure~\ref{fig:desorption}(c), we note a surface 2D TI gap of about 0.17 eV and a DOS shift by 0.05 eV towards $E_{\text{F}}$ compared to the unreconstructed case.
These rather weak changes show that Iodine desorption from the spacer layer only marginally affects the electronic structure at the 2D TI surface already at a slab thickness of 2.5 nm.
The DOS of the spacer surface, Figure~\ref{fig:desorption}(d), however, is completely reconstructed by the desorption:
A gap of about 0.15 eV width is now centred around -0.1 eV and the DOS below the gap is considerably reduced in comparison with the unreconstructed case,
though it is still dominated by Iodine-5p states.
The agreement of this DOS with the STS data~\cite{pauly2} is better than for the case of the previous VCA calculation.
In particular, a relatively high DOS of mixed Bi-6p and I-5p character
is found between the Fermi level and $0.5$ eV. STS data indeed show a hump in this energy region~\cite{pauly2}.
On the other hand, the calculated gap is more narrow than the measured one, which has a width of about 0.3 eV.

Considering the energetics of the modelled process, we find that it requires 2.0 eV per Iodine atom, if desorption of individual atoms is assumed.
Even if we assume, that Iodine molecules are desorbed during the cleavage, an energy of 1.0 eV per Iodine atom is needed.
(Note, that we use for this semi-quantitative discussion our calculated I$_2$ bond energy, 2.0 eV, which is somewhat larger than the experimental value of 1.6 eV.)
This result does not confirm our above hypothesis that the unreconstructed spacer surface might be intrinsically unstable.
Obviously, the applied cleavage~\cite{pauly1} is violent enough to cause an endothermic process.

Summarizing this point, the employed atomistic structure model of the experimentally observed Iodine desorption describes the available spectroscopic information very satisfactorily.
On this basis, we will proceed with the investigation of chemical gating and doping.
All further calculations take into account the described chemical reconstruction of the spacer surface.

  \begin{figure}
      \centering
      \includegraphics[width=0.35\textwidth]{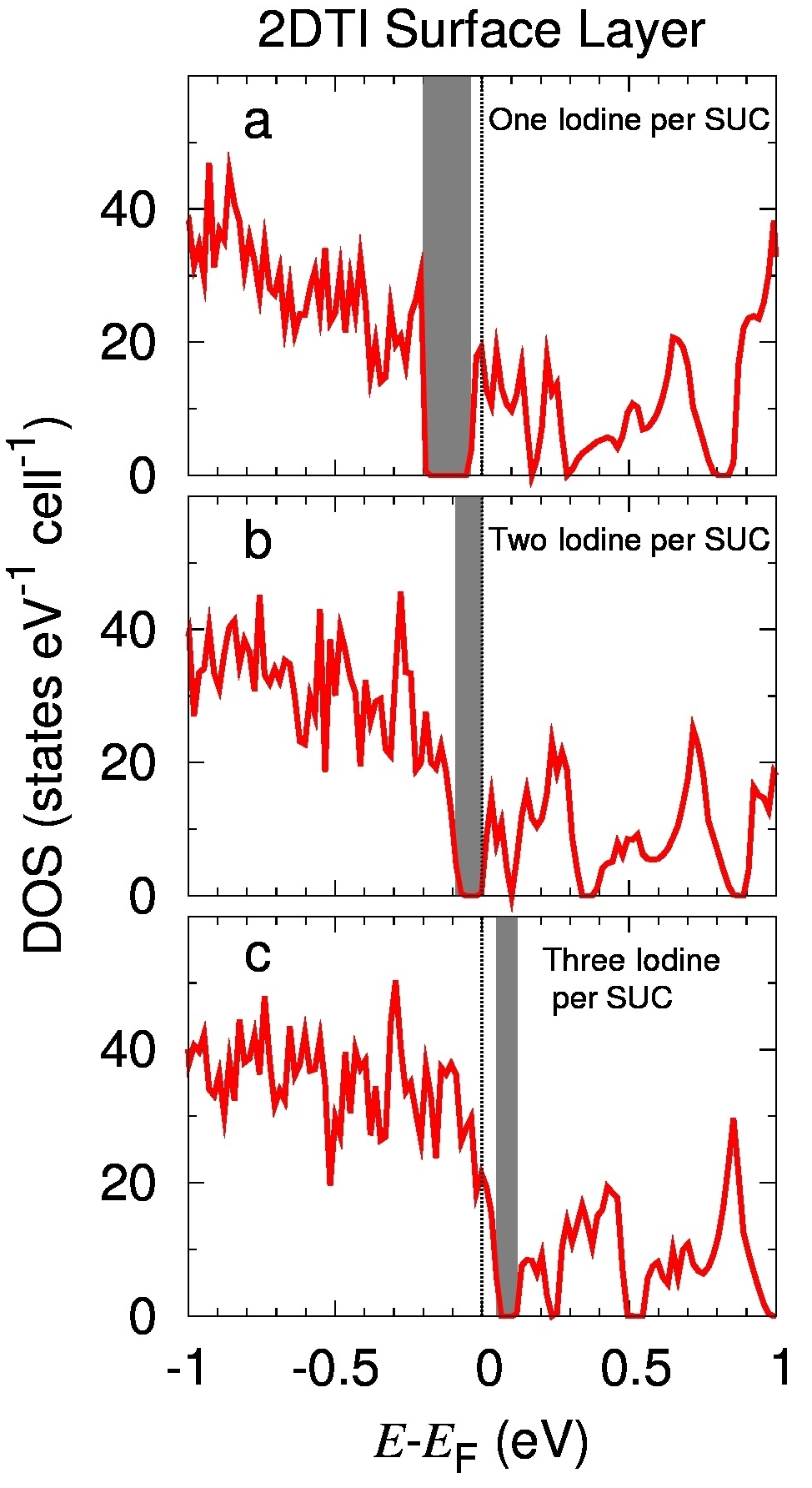}
      \caption{Layer-resolved DOS of the 2D TI surface layer (Bi$_4$Rh)$_3$I$_{1+n}$ for deposition of n=1 (a), n=2 (b), and n=3 (c) Iodine atoms per SUC on top of the 2D TI surface for a chemically reconstructed slab with one structural layer.}
      \label{fig:i123}
  \end{figure}
\textbf{\emph{Adsorption on top of the 2D TI surface:}}
With the aim to compensate the surface charge and to move the 2D TI surface gap with the topological edge states towards $E_{\text{F}}$,
we generate a sparse layer of 1$\ldots$3 Iodine atoms per SUC on top of the 2D TI surface.
This concentration, about 0.08$\ldots$0.25 monolayers, is similar to the concentration of Copper atoms that were recently used in chemical gating of the strong 3D TI Bi$_2$Se$_3$~\cite{wray2013}.
The calculated adsorption energy gain amounts to 1.7 eV for the first 
Iodine atom per SUC, 1.6 eV for the second, and 1.4 eV for the third one, if
deposition of atomized Iodine is assumed. These numbers have to be reduced 
by 1.0 eV for the case of molecular Iodine deposition.

The related DOS contributions of the 2D TI surface are shown in Figure~\ref{fig:i123} for the simplest case of one structural layer.
For the lowest concentration of one Iodine atom per SUC, $E_{\text{F}}$ is shifted downwards in comparison to the pristine case (not shown),
indicating a reduction of electron-type bulk carriers, but stays within the conduction band (Figure~\ref{fig:i123}(a)).
Next, for two Iodine atoms per SUC, $E_{\text{F}}$ moves to the bottom of the conduction band.
The surface band gap of 0.07 eV is smaller than the bulk gap but clearly evident (Figure~\ref{fig:i123}(b)).
Given the position of $E_{\text{F}}$, transport experiments would be feasible.
Further, if three Iodine atoms per SUC are deposited, $E_{\text{F}}$ shifts into the valence band and
a crossover from electron type to hole type behavior occurs (Figure~\ref{fig:i123}(c)).
These findings confirm the naive expectation that the formal surface charge of +2 can be compensated by two Iodine atoms.
In the following, we will restrict our investigation to this adsorbant concentration.

\begin{figure}
      \centering
      \includegraphics[width=0.5\textwidth]{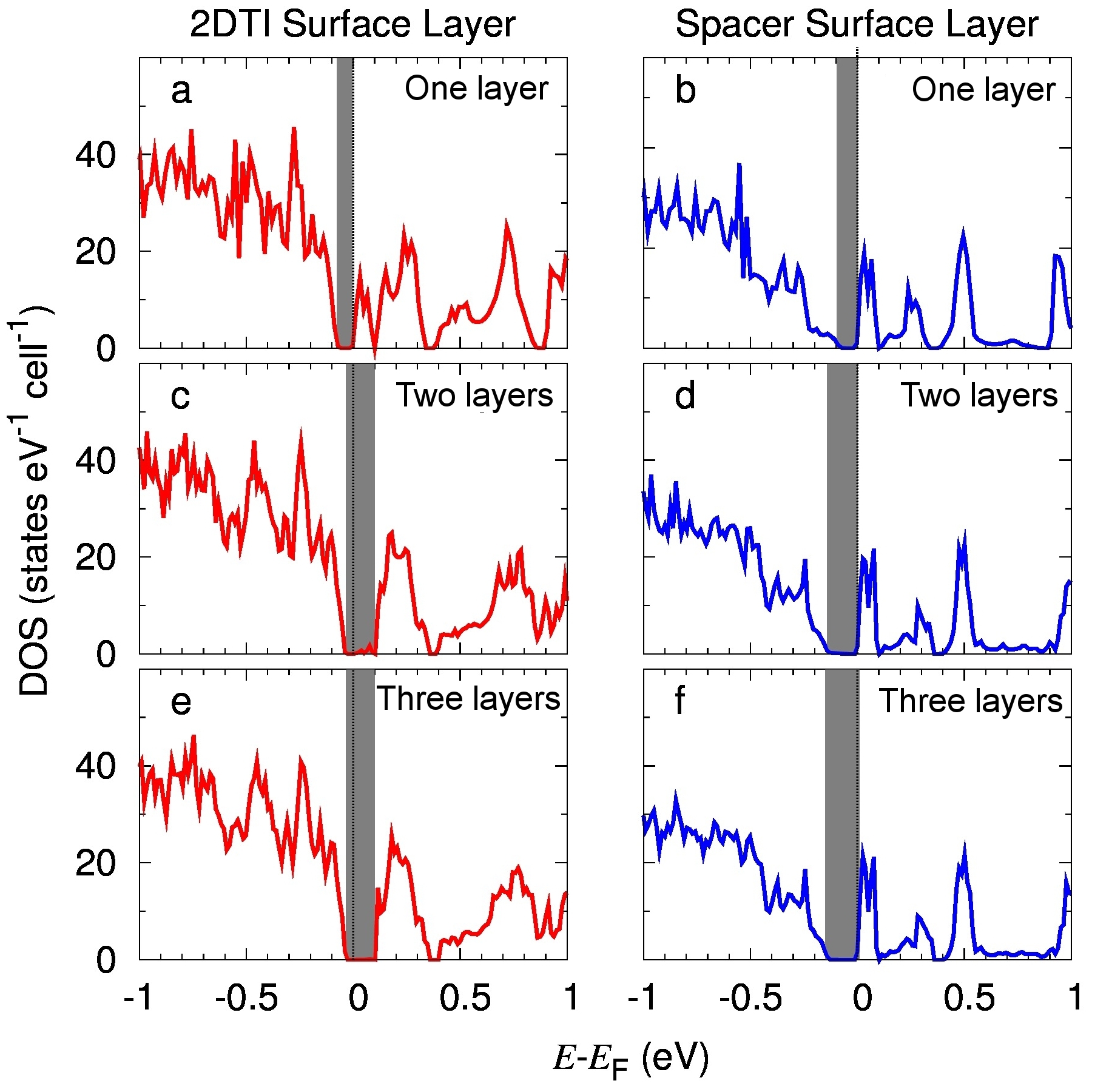}
      \caption{Layer-resolved DOS of 2D TI surface layers (Bi$_4$Rh)$_3$I$_3$ (a, c, e) and spacer surface layers Bi$_2$I$_6$ (b, d, f) for one (a, b), two (c, d) and three (e, f) structural layers with two Iodine atoms per SUC deposited on top of the 2D TI surface. The spacer surface is chemically reconstructed.}
      \label{fig:gating}
 \end{figure}

Figure~\ref{fig:gating} shows the surface-layer projected DOS for slabs of one, two, and three structural layers.
We first consider the 2D TI surface DOS, Figure~\ref{fig:gating}(a, c, e).
In all cases, the valence band and the lower part of the conduction band (up to about 0.35 eV) is dominated by contributions of similar weight 
from Bi-6p and from 5p states located at the adsorbed Iodine atoms. Above the small gap at 0.35 eV, the DOS is dominated by
Bi-6p states. 
With increasing thickness of the slab, the 2D TI gap is found to increase from about 0.07 eV to about 0.12 eV.
The smaller gap size in the case of a slab with only one structural layer is due to the presence of a narrow band with a width of 0.10 eV just above $E_{\text{F}}$.
By comparison with the data of the thicker slabs it becomes obvious that this band originates from hybridisation with spacer surface states being present at the same energy for all considered slab thicknesses, Figure~\ref{fig:gating}(b, d, f).
The integrated weight of that band, projected to the 2D TI surface, amounts to 0.82 (0.05, 0.0034) electrons for slabs with one (two, three) structural layers.
With increasing slab thickness, the interaction between the two surfaces and the related in-gap states at the 2D TI surface becomes weak and finally negligible for slab thickness larger than 2.5 nm.
This means that ultra-thin films, in particular those with only one structural layer, may not be advantageous for the demonstration of the QSH effect in Bi$_{14}$Rh$_3$I$_9$ due to possible narrowing of the gap by interaction with the opposite surface.
Rather, films with a thickness larger than 2.5 nm may serve the goal if their surface is doped with Iodine (or, other oxidizing agents) in an appropriate concentration.
We suggest that the concentration could be naturally stabilized by a self-limited adsorption process, as over-doping might be thermodynamically unstable.
This idea is supported by the calculated adsorption energy gain, which is considerably reduced with growing concentration of adsorbed Iodine.
A fine-tuning of the concentration should be possible by the substrate temperature.

 \begin{figure}
      \centering
        \includegraphics[width=0.75\textwidth]{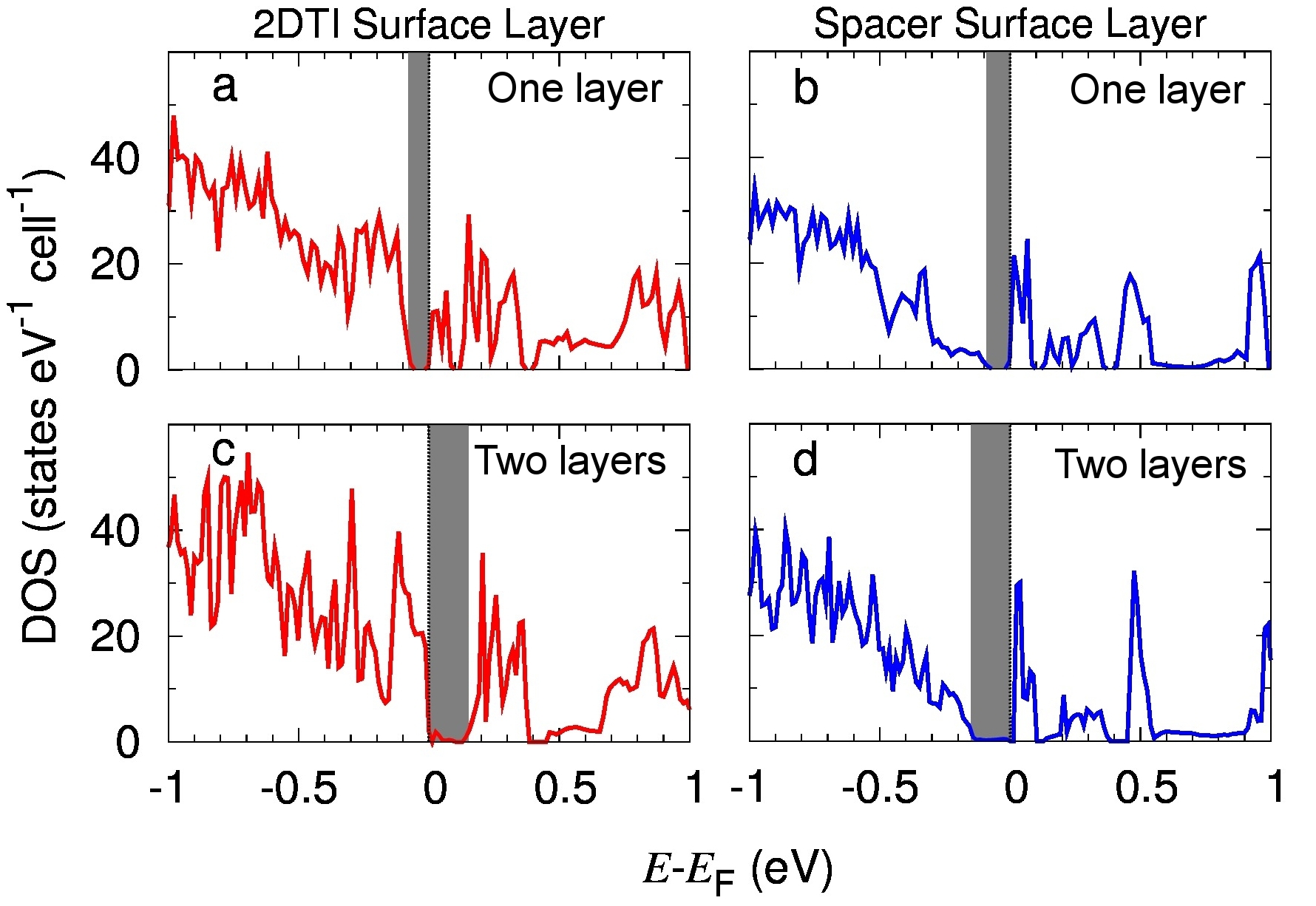}
      \caption{Layer-resolved DOS of 2D TI surface layers Bi$_{10}$Sn$_2$Rh$_3$I (a, c) and spacer surface layers Bi$_2$I$_6$ (b, d) for one (a, b) and two (c, d) structural layers with Sn-doped outermost atomic layer of the 2D TI surface (two Bi atoms replaced by Sn).
              The spacer surface is chemically reconstructed.}
      \label{fig:doping}
 \end{figure}

\textbf{\emph{Chemical doping inside the 2D TI surface:}}
Hole doping through Bi substitution could be another means to move $E_{\text{F}}$ into the 2D TI gap.
This method has been found effective for tuning the carrier type and carrier density of compounds such as Bi$_2$Te$_3$, Bi$_2$Se$_3$ and Bi$_2$Te$_2$Se~\cite{chen,hor,ren1,ren2}.
Here, we consider the partial substitution of Bi atoms in the outermost atomic layer of the 2D TI surface by Sn atoms.
We first investigate the effect of replacing up to three Bi atoms per SUC by Sn for slabs with one structural layer (not shown).
Similar to the case of Iodine deposition, the desired result is obtained, if two nominal charges are removed per SUC.
For this concentration of Sn, the calculated energy cost amounts to 0.7 eV per substituted Bi atom.
As shown in Figure~\ref{fig:doping}, we observe shifting of $E_{\text{F}}$ to the bottom of the conduction band (top of the valence band)
of the 2D TI surface layer in the case of one (two) structural layers.
In both cases, a sizeable surface 2D TI gap of $\sim$0.08 eV ($\sim$0.12 eV) is obtained, similar to the adsorption effect described above.
The character of the electronic states on both sides of the gap is dominated by Bi-6p states, since Sn-5p states show almost the 
same partial DOS per atom as Bi-6p states (not shown).
An effect on the surface spacer is found only in the size of the gap as we move from a slab with one to a slab with two structural layers (see Figure~\ref{fig:doping}).
We note that the considered doping yields very similar results as the chemical gating by adsorption.

\textbf{\emph{Relevance for other materials:}}
Recent STM and ARPES measurements on ZrTe$_5$ suggest this material to be a weak 3D TI \cite{wu,bingli} as well. 
An evidence of topological edge states is observed within an energy gap of 0.1 eV near the step edges on the surfaces \cite{wu}.
However, the non-trivial gap lies $0.03$ eV above $E_{\text{F}}$. This is expected to hinder an observation of the QSH effect as in the case of Bi$_{14}$Rh$_3$I$_9$.
Thus, the present results are relevant for further work on ZrTe$_5$ and other similar weak 3D TI compounds as well.

To summarize, we have demonstrated that chemical gating can compensate the intrinsic n-doping at the surface of Bi$_{14}$Rh$_3$I$_9$, a suggested weak 3D topological insulator.
By deposition of Iodine adatoms in an appropriate concentration or partial exchange of surface Bismuth atoms by Tin, the topological gap is shifted to the Fermi level.
While the former method might be easier implemented for a proof-of-principle experiment, the latter might be more robust for potential applications.
Importantly, the gap is not closed upon chemical gating. As the applied LDA method usually underestimates the gap size, this statement should be robust.
Thus, the gated material will be suitable for transport experiments with the particular aim to confirm its topological character.
We further find that the gap size grows with the thickness of the material.
Therefore, no improvement of the transport-related properties is expected upon extreme reduction of the sample thickness to one structural layer (1.25 nm).
The suggested methods should be of relevance for other, related systems.

\section{Acknowledgements}

M.P.G. thanks the Alexander von Humboldt Foundation for financial support through the Georg Forster Research Fellowship Program.
 We are grateful for discussions with Jeroen van den Brink, Klaus Koepernik, Bertold Rasche, Anna Isaeva, Joseph Dufouleur, Rajyavardhan Ray, and Ion Cosma Fulga.
Technical assistance by Ulrike Nitzsche was very helpful.

\bibliography{refs}


\begin{tocentry}
    \begin{center}
     \includegraphics[width=0.58\textwidth]{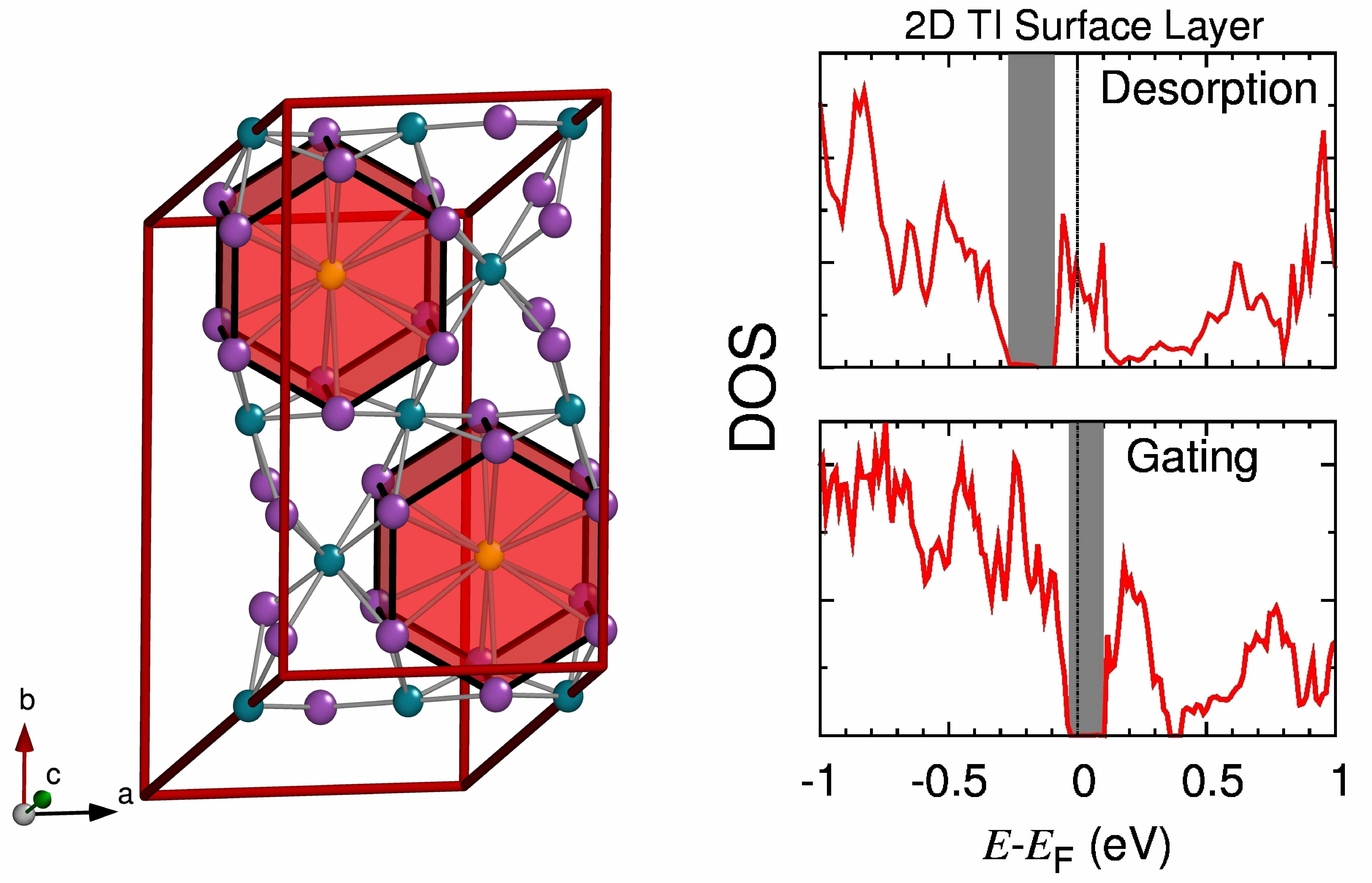}
      \label{for TOC only}
    \end{center}
\end{tocentry}

\end{document}